# Gaussian process analysis of Electron Energy Loss Spectroscopy (EELS) data: parallel reconstruction and kernel control

Sergei V. Kalinin,[1] Andrew R. Lupini,[1] Rama K. Vasudevan,[1] and Maxim Ziatdinov[1]

[1] Center for Nanophase Materials Sciences, Oak Ridge National Laboratory, Oak Ridge, TN 37831

Advances in hyperspectral imaging modes including electron energy loss spectroscopy (EELS) in scanning transmission electron microscopy (STEM) bring forth the challenges of exploratory and subsequently physics-based analysis of multidimensional data sets. The (by now common) multivariate unsupervised linear unmixing methods and their nonlinear analogs generally explore similarities in the energy dimension but ignore correlations in the spatial domain. At the same time, Gaussian process (GP) methods that explicitly incorporate spatial correlations in the form of kernel functions tend to be extremely computationally intensive, while the use of inducing point-based sparse methods often leads to reconstruction artefacts. Here, we suggest and implement a parallel GP method operating on the full spatial domain and reduced representations in the energy domain. In this parallel GP, the information between the components is shared via a common spatial kernel structure while allowing for variability in the relative noise magnitude or image morphology. We explore the role of common spatial structures and kernel constraints on the quality of the reconstruction and suggest an approach for estimating these factors from the experimental data. Application of this method to an example EELS dataset demonstrates that spatial information contained in higher-order components can be reconstructed and spatially localized. This approach can be further applied to other hyperspectral and multimodal imaging modes. The notebooks developed in this manuscript are freely available as part of a GPim package (https://github.com/ziatdinovmax/GPim).

Over the last two decades, scanning transmission electron microscopy (STEM) has become the keystone tool for atomic-level studies of the structure and functionality of solids.[1-2] Structural imaging by STEM now routinely allows locating atomic columns with ~picometer precision[3] and enables the mapping of strain,[4] polarization,[5-10] and ferroelastic[11-13] order parameter fields. Multiple and often spectacular applications of these method for ferroelectric surfaces, interfaces, domain walls, and topological defects have been reported.[12-17]

In parallel, advances in electron energy loss spectroscopy (EELS) opened new pathways for probing materials functionality through energy losses in the electron beam due to inelastic scattering in the material. Core level EEL spectra corresponding to electronic transitions in the solid provide ample information on the presence of specific chemical species, valence states, and orbital populations, although not always in a straightforward manner. This approach has been extensively used to explore single atoms in oxide lattices,[18] charge ordering,[19] oxide interfaces,[20-22] ferroelectric domain walls, etc. A recent surge of interest in monolayer 2D materials has brought a corresponding focus toward EEL spectroscopy of chemical and vibrational[23-25] properties in these systems. Low-loss EELS contains information on the plasmon and exciton excitations and recent advances in monochromated EELS have enabled sub-10 meV resolution, even providing insight into phonons.[23] Recent studies have demonstrated the detection of not only energy loss, but also energy gain due to thermal excitation and laser stimulation.

This remarkable progress in STEM imaging and spectroscopy has necessitated the development of algorithmic tools to denoise/reconstruct the data, extract materials-specific features, and to generally convert the data to materials-specific descriptors that can further feed into atomistic or mesoscopic models. In structural STEM data, typical examples of such analysis are image reconstruction from either high-noise imaging by techniques such as compressed sensing,[26] or from low-noise data by deep learning methods,[27] and identification of atomic positions with associated uncertainty quantification. The former reconstructs images from low-dose or sparse data, whereas the latter converts the image into materials-specific descriptors.

Similarly, analysis of EELS data necessitates the development of corresponding analysis methods. EELS imaging data, by nature, is hyperspectral in that it typically represents the 3D data cube defined by spectra $A(E)$ at some spatial locations $(x,y)$. It is important to note that the EELS signal in STEM is acquired in parallel, with few non-uniform distortions in energy space. In other words, different points in energy are acquired from the same spatial location.

However, analysis of the EELS data cube represents a considerably more complex problem than most structural STEM image data. Similar to many other spectroscopic imaging techniques, analytical or numerical models for EELS signal formation, allowing for all of the instrumental factors, are generally absent or tend to be complicated, creating a need for exploratory data analysis tools. In core-loss EELS, the energy regions corresponding to different atomic species are often localized in energy, allowing for the use of simple peak-fitting tools or even integration across corresponding energy ranges to generate elemental maps. However, this is not always the case. For example, in low-loss EELS, overlap between the peaks corresponding to dissimilar mechanisms are much stronger, again necessitating alternative exploratory data analysis tools.

In our opinion, one of the biggest recent breakthroughs in the analysis of EELS data came with the introduction of unsupervised linear unmixing tools, as envisioned by Bonnett[28-29] and then realized and widely introduced by Kotula and Keenan[30-31] and Watanabe.[32] In this approach, the 3D hyperspectral EELS image is represented as a linear combination of spatially dependent loading maps and energy dependent components, as

$$A_0(x, y, E) = \sum_{i=1}^{N} A_i(x, y) w_i(E) \tag{1}$$

The loading maps, $A_i(x, y)$, represent the variability of the spectral behaviors across the image, and $w_i(E)$ are the components (sometimes referred to as the endmembers) that determine these characteristic behaviors. The number of components, $N$, can be chosen based on the reconstruction error, anticipated physics of the system, etc. Note that Eq. (1) explicitly assumes that the nature of $w_i(E)$ is unknown but that the total response is linear in these components. If the components are known, e.g., if they represent 'pure' spectra, then Eq. (1) will become a linear regression model. The immediate feature of the decomposition is that a $M*L*K$ 3D data set ($M, L$ are the spatial and $K$ the energy dimensions) is reduced to $N \ll K$ spatial maps, each with size $M*L$ and $N$ components of length $K$. For a typical 100x100x1000 EELS data set and $N = 10$, this corresponds to a reduction from $10^7$ data points to $1.1*10^5$ data points, an almost 100-fold compression.

The paradigmatic example of linear unmixing is principal component analysis (PCA),[33-34] in which the components are orthogonal and are ordered by reducing variance. Another example of linear unmixing is non-negative matrix factorization (NMF), where the components are non-negative. Many other unmixing methods are known, for example Bayesian linear unmixing and related methods pioneered by Dobigeon,[35-41] in which the components are both non-negative and sum to one; or independent component analysis (ICA) that aims to maximize non-Gaussianity of the signal. It is important to note that the components of linear unmixing in general do not have direct physical meaning, although in certain cases the constraints such as non-negativity, summing to one, or sparsity allow the user to draw semi-quantitative conclusions using the parallels with the relevant physical mechanisms.

The fundamental limitation of all linear unmixing methods, as well as many of the non-linear manifold-learning techniques, is that they operate in energy space only, whereas spatial correlations in the spatial plane remain unused. In other words, the components in linear unmixing algorithms do not change if the spatial locations ($x,y$) on which they are defined are randomized; this randomization will be reflected in the loading maps only. This deficiency limits the analysis of EELS data and can be expected to affect the reconstruction process.

Here, we explore the applicability of Gaussian process (GP) regression for the analysis and reconstruction of EELS imaging data. Given the large volume of a typical EELS data set, the direct use of a GP method is impractical, requiring either the use of the inducing point approach or similar alternative strategies. The inducing point method often tends to produce reconstruction artefacts, especially for signals with strong gradients (sharp features) that are extremely difficult to detect. To extend the GP methods to hyperspectral data, we develop a kernel transfer approach for dimension-reduced EELS data. We consider two limiting cases, one in which the kernel function is determined by a certain PCA/NMF component and another where the kernel is balanced by several components. We further discuss the reconstruction of EELS data sets using constrained kernels as a way to unify the physics of the signal formation mechanisms. Although we do not discuss this aspect extensively, it is important to bear in mind that the resulting GP methods can also be applied to sparsely sampled data and to cases where some (or even a significant fraction) of the data points are missing. Similarly, once the model is trained the resulting output can be up-sampled or interpolated to predict the expected signal at a higher spatial resolution. Importantly, prior knowledge about the physics or expected mechanisms can be encoded into the kernel. The codes developed here are available as a GPim library on GitHub.

As a model system, we choose the lanthanum aluminate – strontium titanate interface. Data were acquired on a Nion UltraSTEM operated at 100kV and equipped with a Gatan Enfina spectrometer, resulting in a data size of 48x48x1340 pixels (fully described in the Methods).

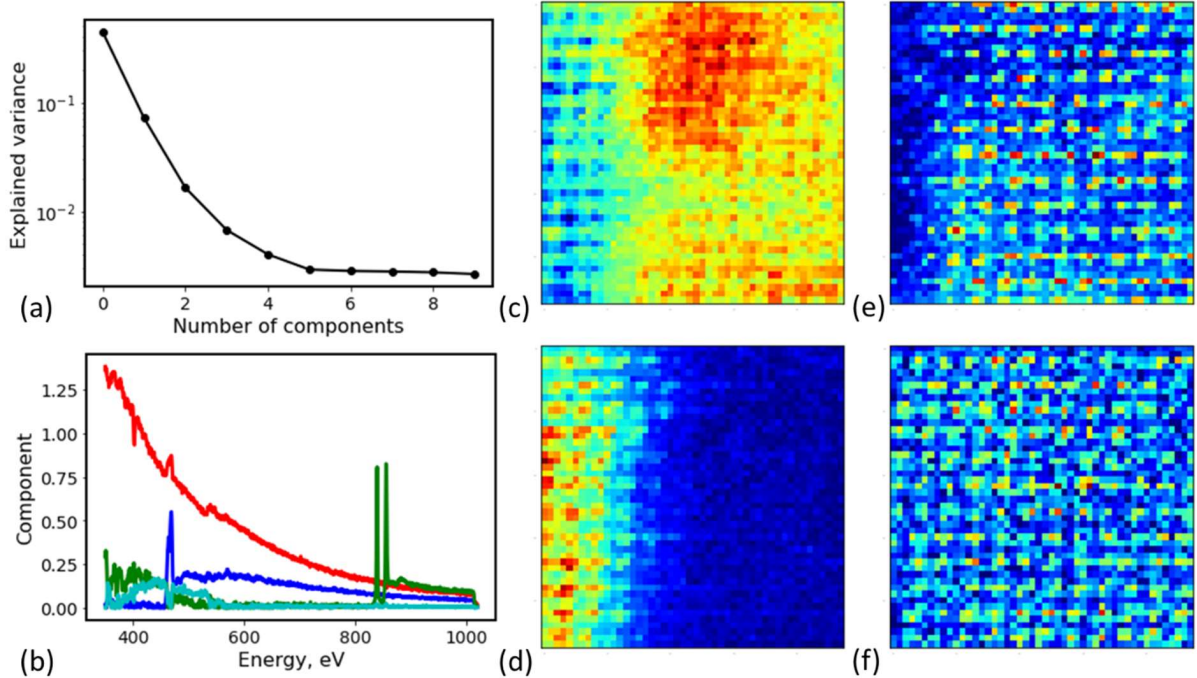

**Figure 1.** (a) PCA scree plot of EELS data set. (b) First four NMF components (red – 1st, blue – 2nd, green – 3rd, and cyan – 4th) and (c-f) first four NMF loading maps.

For pre-processing, a small number of outliers (3 pixels for this data set) were removed using substitution by local averaging. Note that this step is extremely important, since otherwise each outlier can dominate a principal (or NMF) component and result in strong information leakage from other maps. Fig. 1 (a) show the explained variance of the data as a function of the number of components, illustrating that most of the information is concentrated in the first 3-5 components.

To explore the spatial structure of the EELS signal, we adopt a NMF decomposition with $N =12$ components. NMF is chosen here since it allows us to maintain the non-negativity of individual components; however, the GP analysis reported below is universal and can be applied to any decomposition. The first four NMF components are shown in Fig. 1 (b) and the corresponding loading maps are illustrated in Fig. 1 (c-f). Although physical interpretations of NMF components are necessarily qualitative, the first component represents essentially an average signal including the background, the second component corresponds to the signal from the titanium L-edge, the third to the lanthanum M-edge, and the fourth to the oxygen K-edge and some background (component 5 is affected by some afterglow on the spectrometer scintillator and component 6 indicates a difference of the Ti-L edge on and off atomic columns; not shown). Clearly, some atomically resolved features are visible in certain regions for some of the components. Above the 4$^{th}$ component, no atomic-scale features are apparent. In general, atomic features might be expected in all the loading maps (if the corresponding components show peaks corresponding to the core-loss levels); in practice the data is affected by noise and non-optimal sampling.

We explore the reconstruction of the signal using GP regression. This method exploits the presence of correlations within the data set in the spatial domain. A classic GP aims to learn an unknown function, $f$, over source-target pairs, $\{(x_1, y_1), . . .(x_n, y_n)\}$ by performing Bayesian inference in a function space. A standard GP regression model is defined by $f \sim \mathcal{GP}(m(x), K_f(x, x'))$ and $y = f(x) + \varepsilon$, where $K_f$ is a covariance function (usually referred to as a kernel), $m$ is a mean function (usually set to 0), and $\varepsilon$ is Gaussian observation noise. The covariance function determines the strength and functional form of coupling between $y$ values across the parameter space, $x$, and therefore allows, in principle, encoding our prior knowledge into the model. For example, the knowledge of the physics of the system, such as whether or not to expect atomic-scale detail or long-range composition changes, can inform the choice of kernel function.

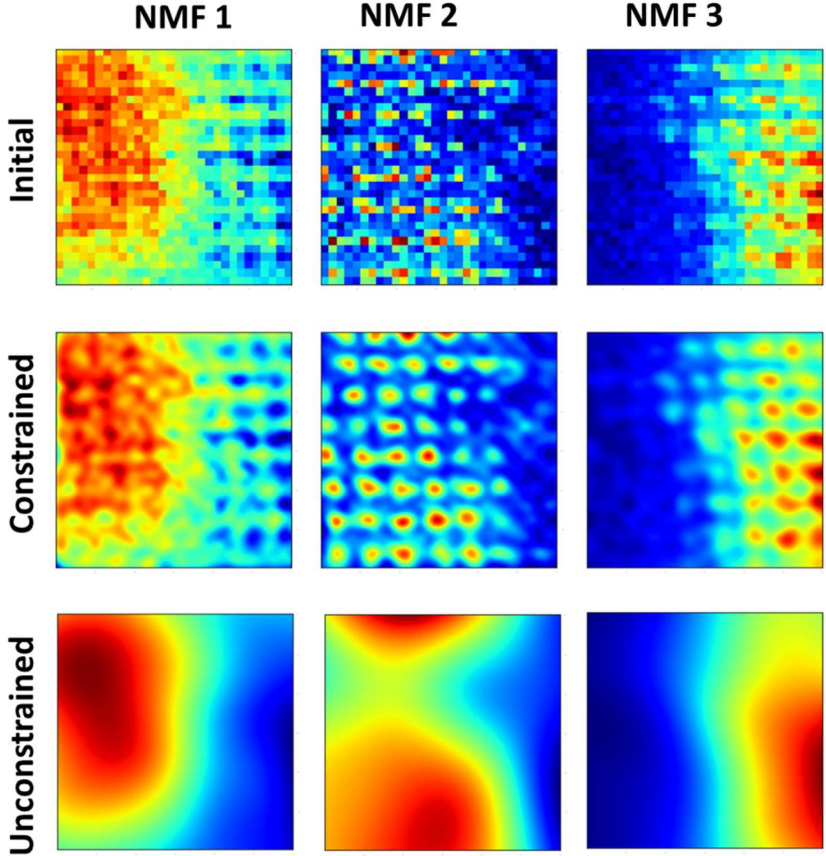

**Figure 2.** Independent GP reconstruction of the first three NMF component maps with constrained and unconstrained kernels. For clarity, analysis is performed on a 30x30 subset of the image. Shown are initial NMF components (top row), constrained GP (higher kernel length limit = 3), and unconstrained reconstructions (bottom row). The resampling is 4 times denser than the initial grid.

The GP reconstruction of the first three NMF components is shown in Fig. 2. Here we used a Matern kernel defined as

$$k_{Matern}(x_1, x_2) = \sigma^2 \exp\left(-\sqrt{5} \times \frac{|x_1-x_2|}{l}\right)\left(1 + \sqrt{5} \times \frac{|x_1-x_2|}{l} + \frac{5}{3} \times \frac{|x_1-x_2|^2}{l^2}\right), \quad (1)$$

where $l(x, y)$ and $\sigma^2$ are kernel length scale and variance, respectively. Note that in our setup the kernel length scale is learned separately in *x* and *y* dimensions (i.e., the kernel is anisotropic). It should also be noted that isotropy and limiting length scales can be imposed as constraints. The convergence of the fit can be explored via the history of the process, namely the evolution of the noise level and the kernel length scale with iterations. Note that in this process, the kernel length scale serves the role of the filter that defines the spatial extent of the features in the image on which the reconstruction converges.

During the analysis, we found that the evolution can proceed in two regimes depending on the chosen kernel constraints. For the constrained kernel, namely GP with an imposed upper limit on kernel length, the GP yields reconstructed images showing both atomically resolved details and large-scale compositional variations, as shown in Fig. 2 (middle row). However, for an unconstrained kernel, the evolution generally proceeds to highlight the large-scale variations in the signal, while the small atomic features are interpreted as noise and smoothed over. This behavior clearly allows an opportunity to separate the physical phenomena via analysis at different length scales, but opens a question as to how to perform this analysis systematically avoiding operator-bias induced artefacts and associated (potentially misleading) interpretations.

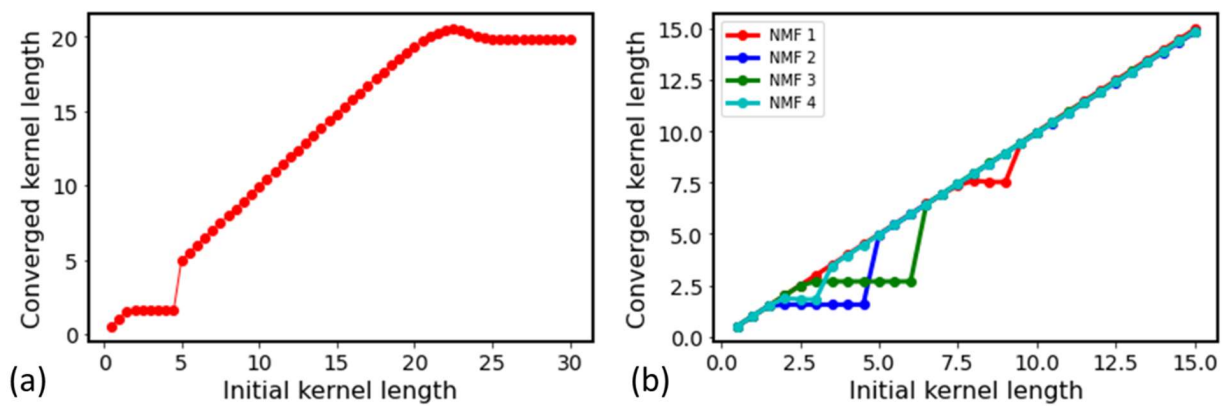

**Figure 3.** Evolution of kernel length in the GP process as a function of limit of kernel length. Shown is (a) behavior for 2$^{nd}$ NMF component and (b) behavior for first four NMF components. For all other NMF components the response yields a straight line.

When exploring the a kernel "size" evolution as a function of the imposed limit, we find that in some cases the evolution approaches the superimposed limit, whereas in others it converges stably to a value corresponding to the characteristic length scale of features in the image. To explore this behavior systematically, we explored the change of the kernel length after GP regression as a function of the limiting kernel length, as shown in Fig. 3. Figure 3 (a) clearly shows that the kernel evolution for the second NMF component has two clear basins of attraction, corresponding to ~2 and ~20. The first of these values corresponds to the size of the atomic features (about 2 pixels) whereas the second represents large-scale variations of contrast due to larger scale effects, such as sample thickness and, of course, the composition variation.

This behavior is further shown for the first four NMF components in Fig. 3 (b). We note that for three of the components, the kernel behavior clearly highlights the length scale of the atomic features and allows us to pinpoint the initial constraint that guides the convergence to this regime. In comparison, all other components show a straight line, indicative of convergence only on the length scales of image inhomogeneities. Overall, the approach described here allows consistent choice of the limiting kernel length scale for the constrained GP reconstruction.

However, the GP analysis illustrated in Fig. 2 reconstructs each NMF map as an independent 2D image, optimizing parameters such as kernel length, amplitude, and noise independently. At the same time, the nature of the NMF components is such that while they represent dissimilar behaviors in the energy dimension, they are defined on the same spatial grid. Correspondingly, the spatial correlations within the maps can be expected to be similar, necessitating transfer of information between components during the GP analysis.

We implement a version of GP for vector valued functions with a common spatial structure (i.e., multiple outputs sharing the same inputs). In this case, the covariance can be defined as $k([x,l],[x',l']) = k_l(l,l')k_x(x,x')$, where $k_l$ and $k_x$ represent the correlation between outputs and a standard covariance function operating on inputs, respectively.[42] The former is expressed as $k(l,l') = (BB^T + diag(\mathbf{w}))_{l,l'}$ where $B$ is a low-rank matrix and $\mathbf{w}$ is a non-negative vector. These hyperparameters are trained together with the hyperparameters of the input covariance function, using marginal log likelihood as a "loss" function. Here, each output is associated with a different effective noise, $\varepsilon_l$, which is the GP model's hyperparameter and is also learned during the training. The trained GP model can then be used to calculate the predictive mean and variance on the new data points in the same way as a standard scalar GP.

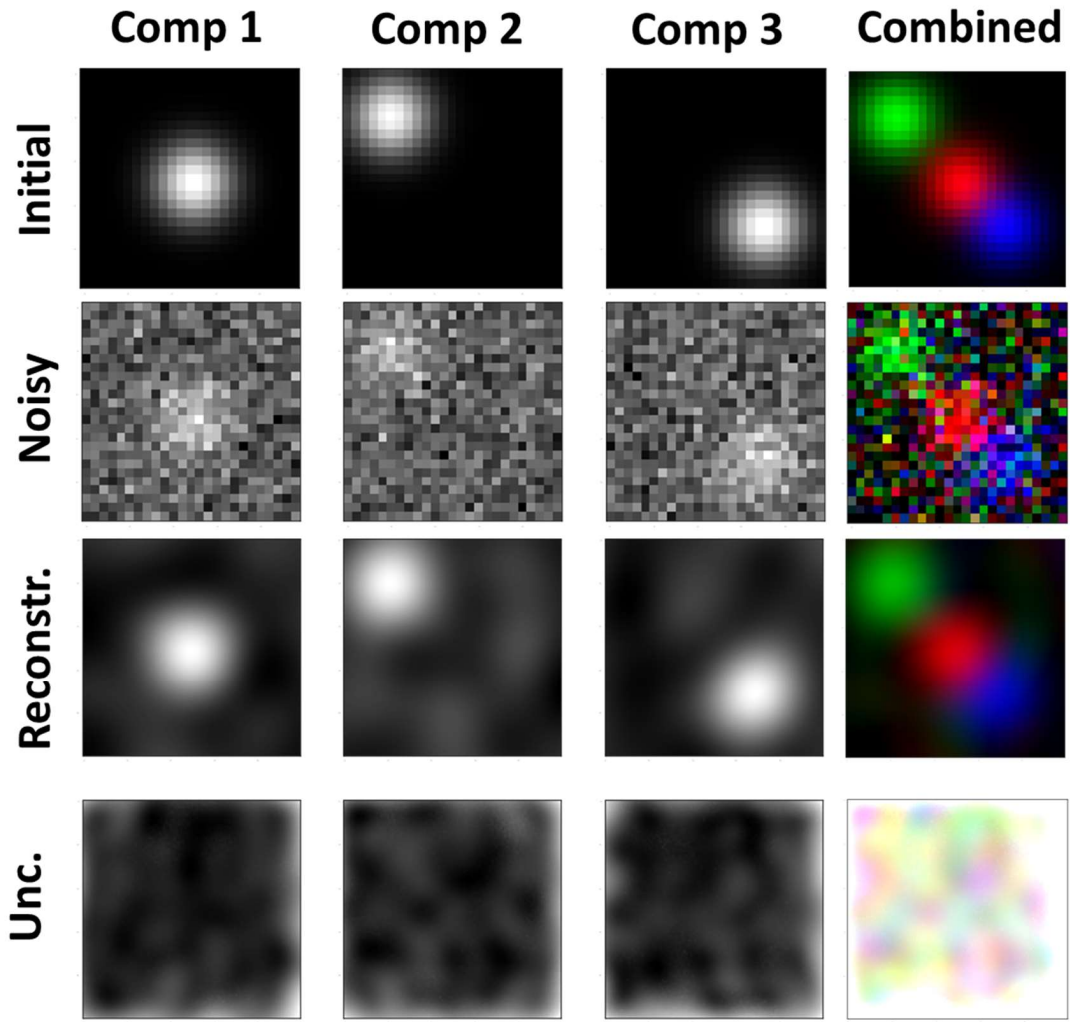

**Figure 4.** Parallel GP reconstruction of the 3-component data set. Shown is the ground truth data (top row), noise corrupted data for noise level $\sigma = 0.3$ (second row), reconstructed data resampled at 4 times the original grid density (third row), and reconstruction uncertainty (bottom row). The scale for all figures is [0,1]. For combined images, each RGB component is displayed in its own scale [0,1].

To illustrate this approach and to better see how to apply it to our experimental data, we first explore a synthetic data set, as shown in Fig. 4. We consider a signal comprised of three components, shown in the top row of Fig. 4. For convenience and compactness of illustration, these components can be represented as a red-green-blue (RGB) image, efficiently encoding the information and allowing for easy interpretation (last column). For this example, the contrast varies from 0 to 1 and the vertical scale of the images is correspondingly normalized. The second row represents the data with the addition of uncorrelated (spatially) Gaussian noise with magnitude $\sigma$ = 0.3. The third and fourth rows represent the GP reconstruction and the associated uncertainties respectively.

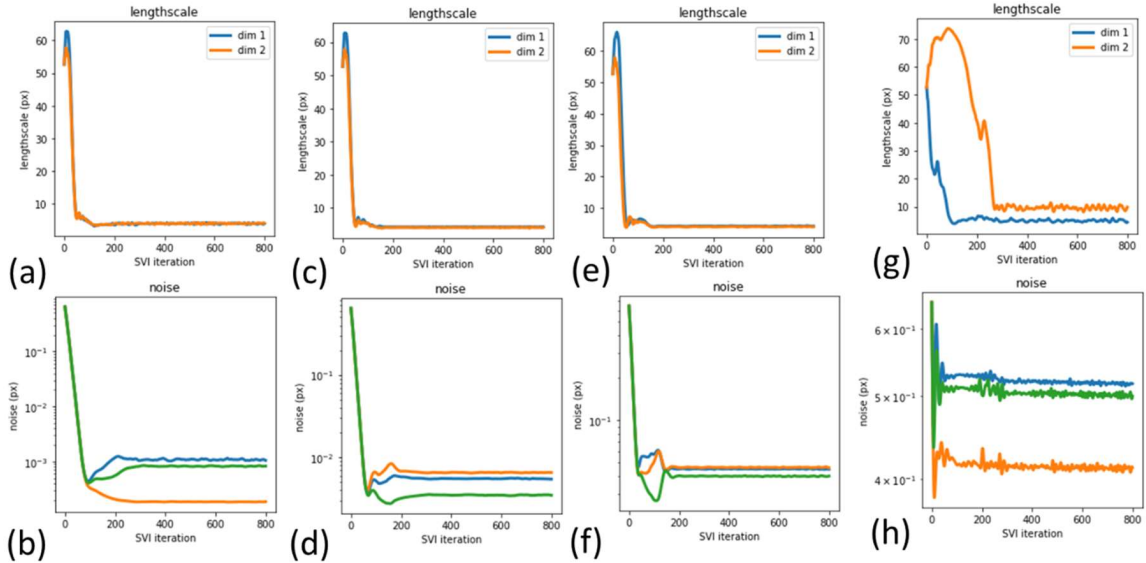

**Figure 5.** Evolution of the parallel GP training on the 3-component synthetic data set in Fig. 4 for different noise levels with anisotropic unconstrained kernel. Shown (a,c,e,g) is kernel length scale evolution and (b,d,f,h) noise evolution. The GP is performed for noise levels (a,b) $\sigma$ = 0.03, (c,d) $\sigma$ = 0.1, (e,f) $\sigma$ = 0.3, and (g,h) $\sigma$ = 1. 1.

The corresponding training histories are shown in Fig. 5 along with the evolution of the kernel length scales and effective noise during parallel GP reconstruction. Note that here the kernel is anisotropic 2D, describing the spatial correlations within the image planes. The kernel is common between the three images, while the noise levels are independent. In all cases, in the initial stages of GP reconstruction, the effective kernel length scale increases and the noise rapidly decreases as the algorithm aims to establish the length scale of correlations in the multimodal image. After this initial stage, the length scale starts to decrease and eventually stabilize and the noise also stabilizes. It is important to note that the kernel length scale is determined by the correlations present in the image, but is not necessarily the best measure of the feature size. For low noise levels, the kernel lengths are similar, whereas for the high noise levels, the lengths tend to split during reconstruction. For very high noise levels (not shown) the kernel length can demonstrate even more complex dynamics, with one length saturated and another oscillating with time. These behaviors, which quite clearly indicate where the model is unsuccessful, can be used to establish the stability of the reconstruction process.

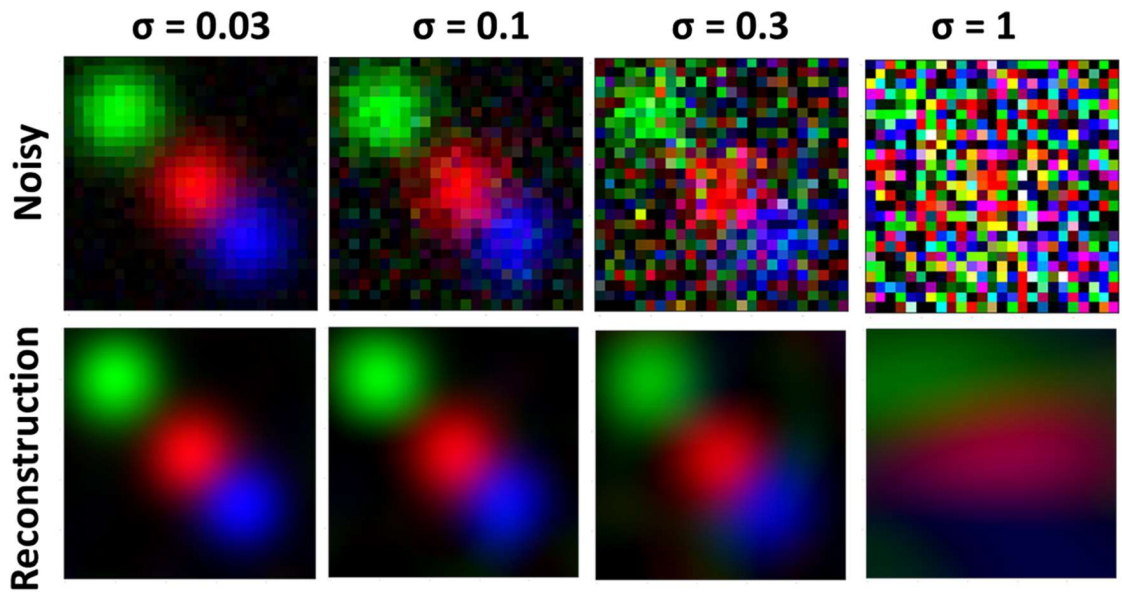

**Figure 6.** Data (top row) and parallel GP reconstruction (bottom row) for several noise levels. The vertical scale is [0,1] for all RGB components. The reconstructed images are resampled on a 4 times denser grid.

To obtain insight into the quality of the reconstructions, Fig. 6 shows the reconstruction with an unconstrained anisotropic Matern kernel for the synthetic data as a function of the noise level. Here, we use an RGB representation of the three component ground truth images in the same manner as in Fig. 4. This representation allows us to both conveniently visualize the data set and to determine the relative changes between the components. For example, if all three components are maximum, the pixel is white; for all three being zero, the pixel is black; and if only one component is non-zero the pixel has one of the primary red, green, or blue colors depending on which component it is and if several components are non-zero a mixed color is seen. Visual inspection of Fig. 6 shows that the features are reconstructed with high veracity up to a noise level $\sigma = 0.3$, whereas for $\sigma = 1$ the reconstruction is clearly degraded. That said, it is important to note that the presence and positions of the features can be established by the GP even for these high noise levels, whereas visual inspection of the unreconstructed image barely reveals any spatial features (right-most column of Fig. 6). Hence, we conclude that while the human eye offers generally a good guide to the presence of noisy features in the image, the GP algorithm might be expected to perform at an even higher noise level than human perception.

      These analyses suggest that the GP algorithm can potentially allow reconstruction at better than human detection levels, that limiting the kernel lengths plays an important role in the reconstruction process as a regularizing factor, and that the parallel GP method allows for information transfer between components of multimodal images in the form of (isotropic or anisotropic) kernel length. Below, we explore the salient features of this parallel GP process, seeking to answer the question: to what extent does the knowledge (i.e., low-noise level) of one component allow us to improve the reconstruction of other components? How is this process affected by kernel constraints? And will reconstruction of the low-noise (well known) component be affected by the presence of the high-noise components?

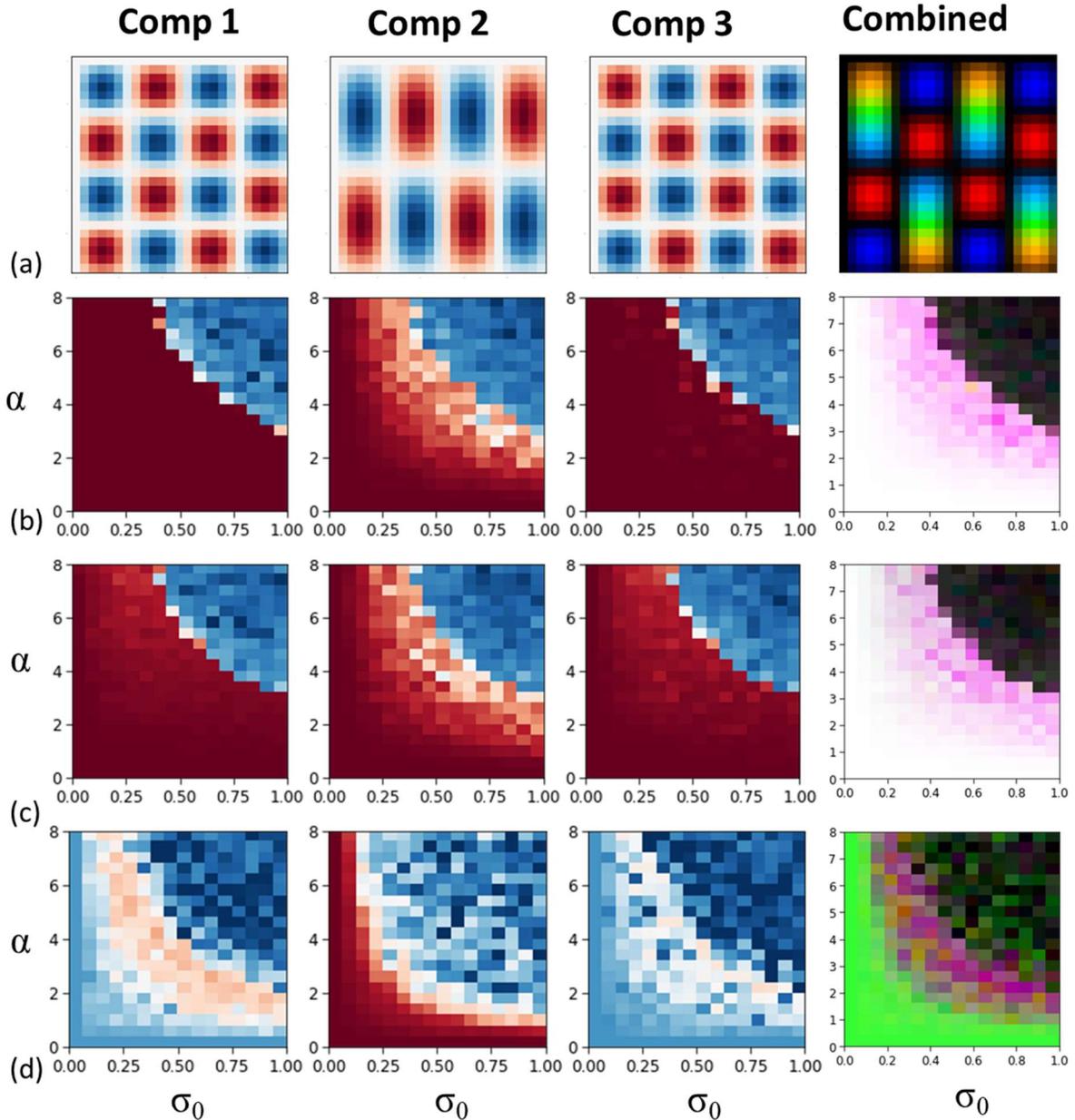

**Figure 7.** (a) Synthetic 3-component data set in the single component and RGB representation. (b-d) Similarity between the ground truth and reconstructed data for all three components in absolute and RGB representation as a function of noise level (horizontal) and noise ratio (vertical). Shown is the analysis for (b) fully known component 1 and free kernel, (c) partially known component 1 and free kernel, and (d) incorrectly constrained kernel.

To explore these questions systematically, we introduce a different ground truth data for the three components, as shown in Fig. 7, using the product of sine functions. Here, components 1 and 3 are identical and periodic, whereas component 2 has similar periodicity in one direction and double the periodicity in the orthogonal direction. This choice of synthetic data set is driven by the obvious parallel with the EELS problem, where atomically resolved features are visible on some of the NMF maps but not on others and there are potential pitfalls for the reconstructions, such as period doubling, but in general the signal is expected to have periodicity commensurate with the underpinning lattice. We note that, as for any synthetic data set, optimization and assessment of performance of the algorithm for each specific problem necessitates a synthetic data set that captures the salient features of the relevant physics.

To quantify the performance of the reconstruction process, we introduce the similarity, $sim_i$, of the noiseless ground truth image for the $i$-th component and the corresponding GP reconstructed image as a simple cross-correlation between the two. If the reconstructed image is identical to the ground truth image, $sim_i = 1$, the reconstruction is ideal and if $sim_i \ll 1$, the reconstruction fails. The similarity function is defined for all three components and can also be represented in an RGB format. The RGB representation allows easy detection of components that start to degrade first with increased noise level based on the hue. Obviously, this analysis is possible only when the ground truth image is known (as is here) or postulated in some manner.

We further create noisy data sets where each component 1-3 is corrupted by uncorrelated Gaussian noise. To better explore the properties of the reconstruction, we use several different levels of noise across the components. In model 1, noise magnitudes are taken as $\sigma_1, \sigma_2, \sigma_3 = (0, \alpha \sigma_0, \alpha \sigma_0)$, where $\alpha$ is a scaling factor and $\sigma_0$ is an absolute noise level. In model 2, noise magnitudes are taken as $\sigma_1, \sigma_2, \sigma_3 = (\sigma_0, \alpha \sigma_0, \alpha \sigma_0)$. Thus, model 1 allows us to explore to which extent the presence of the noiseless image (component 1) in parallel GP affects the reconstruction of noisy images with dissimilar (component 2) and identical (component 3) spatial structures. Model 2 allows us to access the effect of noise in the first component, a necessary comparison given that coupling between kernels is based on the covariance matrix, which is affected by noise in the system. The similarity is then plotted as a function of $\alpha$ and $\sigma_0$, $sim_i(\alpha,\sigma_0)$. Note that while this representation is, strictly speaking, redundant, it allows for easier interpretation of the resulting dependencies and yields insight into the reproducibility of the reconstruction.

The similarity analysis for the unconstrained kernel and model 1 (i.e., noiseless component 1) is shown in Fig. 7 (b). Here, the $\sigma_0$ (horizontal axis) was varied from 0 to 1 and $\alpha$ (vertical axis) was varied from 0 to 8. Thus, the left and bottom lines represent zero noise reconstructions, and the top right corner represents the reconstruction when the noise level is 8 times the maximal contrast.

The behavior of the *sim*$_1$ component suggests that for low noise levels the zero-noise image can be reconstructed very well. However, for sufficiently large noise levels on the 2$^{nd}$ and 3$^{rd}$ components the reconstruction fails, since the kernel attempts to share the information between all three components equally. The reconstruction failure in this case is very sharp, as evidenced by abrupt transitions the between red (*sim*$_1$ = 1) and blue (*sim*$_1$ = 0) regions in Fig. 7 (b). For the 2$^{nd}$ (doubled) component the transition between the good and bad reconstruction is more gradual. Examination of the spatial maps (not shown) in this case suggests that while some spatial features are reconstructed, the others can be shifted, resulting in only a partial overlap between ground truth and the reconstructed image. Finally, an interesting behavior is observed for the third component where the ground truth image is identical to component 1. In this case, the high-quality reconstruction areas for *sim*$_1$ and *sim*$_3$ are almost identical, despite the presence of non-unity pixels in *sim*$_1$. This behavior is further depicted as an RGB map, where the extent of the purple (red for component 1 and blue for component 3) region depicts the extent of improved reconstruction of the 1$^{st}$ and 3$^{rd}$ components compared to the 2$^{nd}$. These observations suggest that parallel GP improves the quality of the reconstruction when the spatial structure of the images is similar.

This result is useful because it illustrates how to apply parallel GP to EELS data: we would expect the parallel GP to provide a benefit when different components share a similar localization or ordering (we might expect some core-losses to be localized near to the corresponding atomic columns and so on).

The reconstruction for model 2 is illustrated in Figure 7 (c). Here, it is clearly seen that the presence of noise in component 1 affects the reconstructions of the three components differently. For component 1, we observe the effect of noise leakage from components 2 and 3 as a gradual decay of the reconstruction quality in vertical direction (remember that the noise for three components is $\sigma_1, \sigma_2, \sigma_3 = (\sigma_0, \alpha\, \sigma_0, \alpha\, \sigma_0)$). However, the transition between red and blue regions is still sharp. For the second component, the reconstruction quality changes weakly and similarity maps *sim*$_2(\alpha,\sigma_0)$ look almost similar for models 1 and 2. Finally, for the third component the behavior is almost similar to the first. This behavior suggests that the reconstruction of two components with identical spatial structure and different noise level balances through the kernel, i.e., they behave like a single image. This effect does not extend to an image with different spatial features.

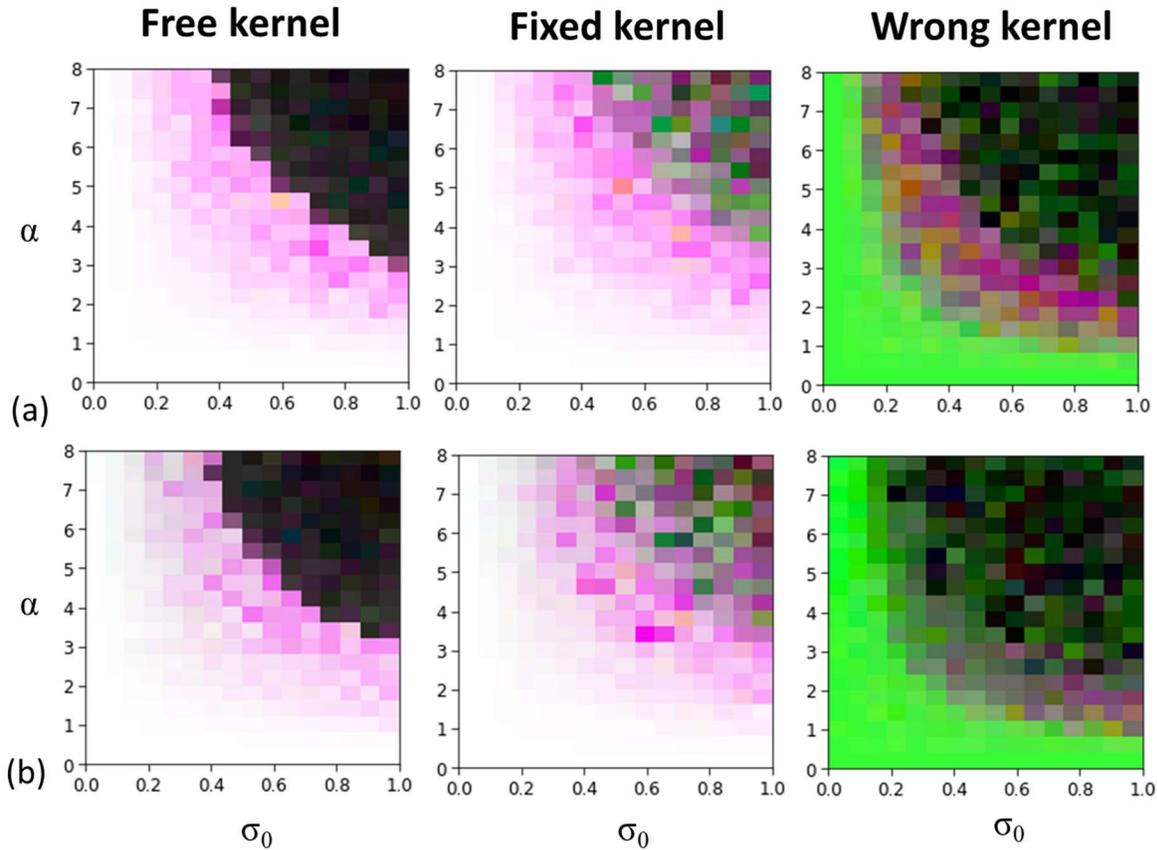

**Figure 8.** Effect of kernel constraint on reconstruction. Shown are results for (a) model 1 and (b) model 2 for free kernel (left column), kernel constrained around the characteristic length scale (central column), and kernel constrained around the value twice larger than characteristic length scale.

Finally, we explore the effect of kernel constraints on the reconstruction, as shown in Figure 8. The behavior in the left column represents the free kernel for models 1 and 2 and are identical to those in Fig. 7. In comparison, the second column illustrates the behavior of the kernel constrained as shown in Figure 7 (d) and is the reconstruction where the kernel is constrained to the [2, 5] interval, close to the value of ~4.5 for the reconstruction of zero noise data (i.e., intrinsic kernel length for this data set). The effect of the optimal kernel on the reconstruction is immediately obvious as the reduction of the dark region in the top right corner of the diagrams. Hence, reconstruction become possible at much higher noise levels if the kernel interval is known correctly. However, if the wrong kernel length is chosen corresponding to an incorrect assumption on the physics of the system, the reconstruction fails completely, as shown in right column for a kernel confined to the range [10, 11] pixels.

This behavior for individual components is shown in Fig. 7 (d). Note that the reconstruction converged only for the $2^{nd}$ component (since the features are twice as large), but failed for the $1^{st}$ and $3^{rd}$ component even for low noise levels. Interestingly, the reconstruction is partially successful for intermediate noise levels where the GP algorithm has sufficient flexibility to discover the extant features despite the deliberate attempt to impose a faulty model.

These analyses suggest that the parallel GP method proposed here can be a powerful paradigm for the reconstruction of multimodal imaging data with a common spatial support and varying noise levels. The quality of the reconstruction can be improved significantly if the kernel length scale is known; however, the incorrect choice of kernel usually leads to the failure of the reconstruction.

We note that the *a priori* length scale for kernel reconstruction is unknown. However, we propose to use the analysis shown in Fig. 3 to derive the relevant kernel length scale. In other words, we use the kernel convergence intervals determined for low-noise components to impose a joint constraint on all components in the analysis. This approach for the NMF loading maps is illustrated in Figure 9, where the kernel interval is chosen to be [0, 2.5] pixels.

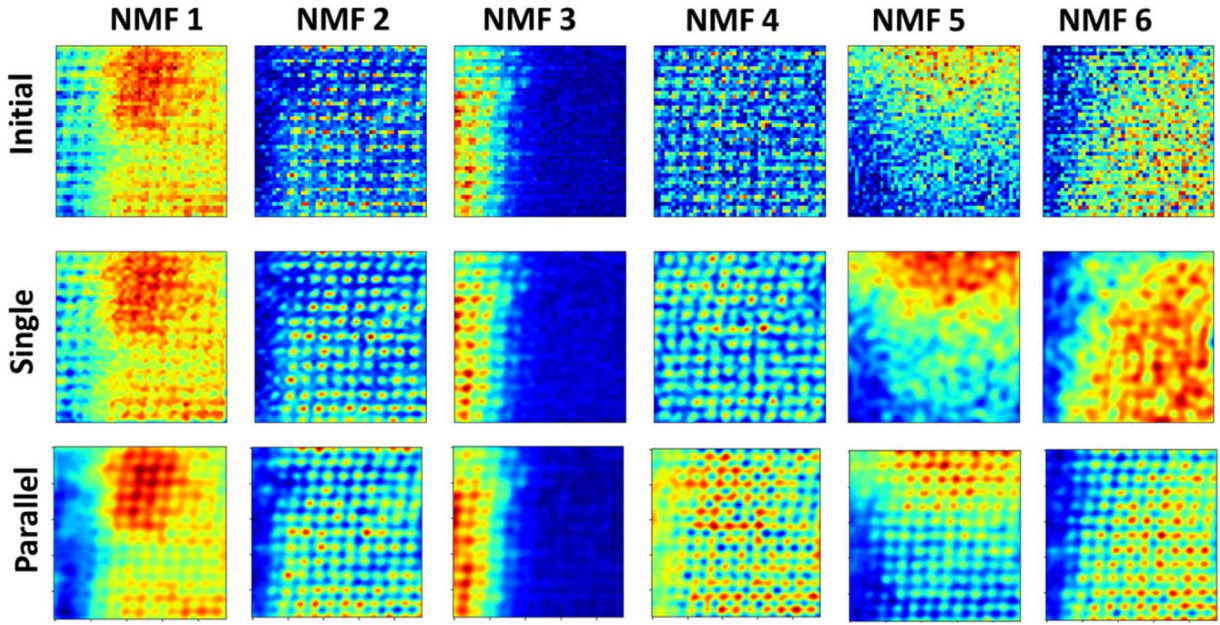

**Figure 9.** Comparison of individual and parallel GP reconstruction of \NMF loading maps. Shown are (top row) original NMF components (similar to Fig 1) and (middle row) individual GP reconstructions. Shown in the bottom row are parallel GP reconstruction on all 6 components simultaneously. Kernel constraints used were the same for the individual and parallel GP. Note that the full 48x48 spatial pixel images are analyzed but due to memory constraints, resampling for parallel GP reconstruction is by a factor of 2. Corresponding spectral components remain unchanged.

The reconstruction of the NMF data set on the full spatial grid is shown in Fig. 9. It is clearly seen that the GP reconstruction of the individual components (even with kernel constraints) yields atomic-scale contrast for the first four components and fails for component 5 and 6. On the other hand, parallel GP clearly allows us to reconstruct the atomic-scale features in these components. However, analysis of a larger number of components does not lead to further improvement. Using 7 components leads to partial degradation of contrast and then a full loss of atomic periodicities for 8 components (not shown). This reveals that the model is effectively using knowledge from the lower noise components in the reconstruction of the weaker signals.

To summarize, we explored the applicability of Gaussian process (GP) methods for the analysis and reconstruction of EELS data sets in STEM. The typical data volumes in this method make direct high-dimensional GP impractical while the use of the inducing point method tends to corrupt the fine features in the energy and spatial dimensions. We therefore suggest and implement the parallel GP method operating on the full spatial domain and a reduced representation in energy domains obtained via linear unmixing. In this parallel GP, the information between the components is shared via a common kernel structure while allowing for variability in relative noise magnitude or image morphology. Note that unlike methods such as transfer learning in convolutional neural nets, the kernel for multiple images here is learned jointly rather than relying on the previous parameters.

Using synthetic data that emulates some characteristic aspects of atomic-resolution EELS data sets, we demonstrate that this approach significantly improves the quality of the reconstruction. We further show that kernel constraints also allow us to increase the quality of the reconstruction and we suggest an approach for estimating these from the experimental data based on kernel length scale convergence analysis for individual components.

Application of this method to EELS data sets demonstrate that spatial information contained in higher-order components can be reconstructed and spatially localized. We believe that this method can be further applied to other hyperspectral and multimodal imaging modes where the data volumes preclude direct application of multidimensional GP reconstructions. The notebooks developed in this manuscript are freely available as a part of the GPim package (https://github.com/ziatdinovmax/GPim).

**Competing Interests:**

The Authors declare no Competing Financial or Non-Financial Interests.

**Acknowledgements:**

This effort (electron microscopy, Gaussian Process workflow) is based upon work supported by the U.S. Department of Energy (DOE), Office of Science, Basic Energy Sciences (BES), Materials Sciences and Engineering Division (S.V.K., A.R.L.) and was performed and partially supported (GPim development by M.Z. , R.K.V.) at Oak Ridge National Laboratory's Center for Nanophase Materials Sciences (CNMS), a U.S. Department of Energy, Office of Science User Facility.

**Author Contribution:**

S.V.K. proposed the concept and led the paper writing. A.R.L. obtained STEM/EELS data. M.Z. wrote the code for Gaussian processes-based data reconstruction and analysis. S.V.K analyzed synthetic and experimental data. R.K.V. assisted in interpretation of results. All authors contributed to paper writing.

**Materials and methods:**

Data were acquired on a Nion UltraSTEM 60-100 operating at 100 kV and equipped with a Gatan Enfina spectrometer with nominal convergence and collection angles of 30 and 48 mrad and a high-angle annular dark field (HAADF) detector inner angle of 86 mrad with an exposure time of 0.1 s/pixel and a dispersion of 0.5 eV/channel. The size of the resulting data set is 48x48x1340, with approximately 0.1 nm/pixel spacing between probe positions, and a 16x16 sub-scan used at each point. The samples were rather challenging as they tended to exhibit either charging or contamination at the relevant interfaces. A small amount of drift and some sample charging caused distortion across the scan, giving a resulting field of view of about 4.4 x 4.4 nm. The survey image is shown below.

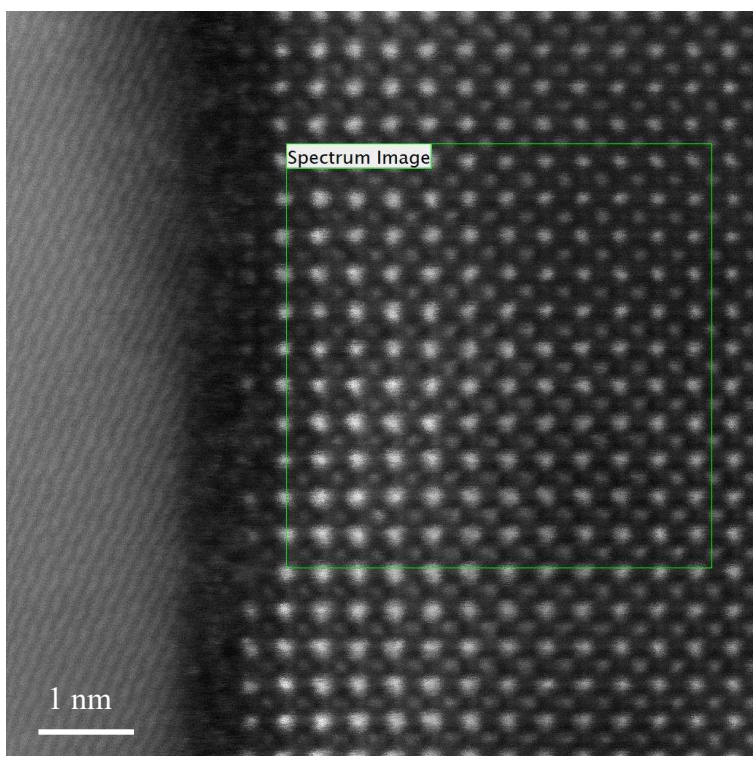

**Figure 10.** Survey image of the sample used for acquiring EELS data.

**Data availability:**

The code is available at https://github.com/ziatdinovmax/GPim. The experimental data is available from the authors upon reasonable request.